\documentclass[11pt]{article}

\usepackage{a4wide}
\usepackage{times}
\usepackage{amsfonts}
\usepackage{amsmath}
\usepackage[psamsfonts]{amssymb}
\usepackage{latexsym}
\usepackage{color}
\usepackage{graphics}
\usepackage{enumerate}
\usepackage{amstext}
\usepackage{url}
\usepackage{epsfig}

\newcommand{\cX}{\cal X}

\newcommand{\Z}{{\mathbb Z}}
\newcommand{\N}{{\mathbb N}}

\newcommand\independent{\protect\mathpalette{\protect\independenT}{\perp}}
\def\independenT#1#2{\mathrel{\rlap{$#1#2$}\mkern2mu{#1#2}}}

\newtheorem{Theorem}{Theorem}
\newtheorem{Def}{Definition}
\newtheorem{Lem}{Lemma}

\title{Causal Markov condition for submodular information measures} 

\author{
Bastian Steudel$^1$, Dominik Janzing$^2$ and Bernhard Sch\"olkopf$^2$\\
${}$\\
{\small 1) Max Planck Institute for Mathematics in the Sciences}\\
{\small Leipzig, Germany}\\
{\small 2) Max Planck Institute for Biological Cybernetics} \\
{\small T\"ubingen, Germany}
${}$\\
}

\date{February 19, 2010}

\begin{document}

\maketitle

\begin{abstract}
The causal Markov condition (CMC) is a postulate that links observations
to causality. It describes the conditional independences among the
observations that are entailed by a causal hypothesis in terms of a
directed acyclic graph.
In the conventional setting,  the observations  are random variables and
the independence is a  statistical one, i.e., the information content 
of observations is measured  in terms of Shannon entropy. 
We formulate a generalized CMC for any kind of observations on which
independence is defined via an arbitrary submodular information measure.
Recently, this has been discussed for observations in terms of binary
strings where information is understood in the sense of Kolmogorov
complexity. Our approach enables us to find computable
alternatives to Kolmogorov complexity, e.g., the length of a text after
applying existing data compression schemes. We show that our CMC is
justified if one restricts the attention to a class of causal mechanisms
that is adapted to the respective information measure.
Our justification is  similar to  deriving  the statistical CMC from
functional models of causality, where every variable is a deterministic
function of its observed causes and an unobserved noise term.

Our experiments on real data demonstrate the performance of compression based causal inference.
\end{abstract}

\section{Introduction}

Explaining observations in the sense of inferring the underlying causal structure
is among the most important challenges of scientific reasoning.
In practical applications it is generally accepted that causal conclusions
can be drawn from observing the influence of interventions. 
The more challenging task,  however, is to infer causal relations on the basis of non-interventional observations and research in this  direction still is considered with
skepticism. It is therefore important to thoroughly formalize
the assumptions and discuss the conditions under which they are satisfied.
For causal reasoning from statistical data,
Spirtes, Glymour, Scheines~\cite{spirtes} and Pearl~\cite{pearlCausality} formalized the assumptions under which
the task is solvable. With respect to a causal hypothesis in terms of a directed acyclic graph (DAG) the most basic assumption is the causal Markov condition
stating that every variable is conditionally independent of its non-descendants, given its parents,
\[
x_j \independent nd_j \,| pa_j\,,
\]
for short.
Pearl argues that this follows from a ``functional model'' of causality
(or non-linear structure equations), where every  node is a deterministic function of  its parents 
$pa_j$ 
and an unobserved noise term $n_j$ (see Fig.~\ref{figFunc}), i.e.,
\begin{equation}\label{func}
x_j=f_j(pa_j,n_j) \,.
\end{equation}
The causal Markov condition is then a consequence of the statistical independence of the noise terms, which is called \emph{causal sufficiency}. It can be justified by the assumption that every dependence between them requires a common cause (as postulated by Reichenbach~\cite{Reichenbach}), which should then explicitly appear in the causal model.  From a more abstract point of view, condition~(\ref{func}) can be interpreted
as saying that the node $x_j$ does not add any more information that is not already contained in
the parents and the noise together. If  we restrict the 
assumption to discrete variables, the corresponding information measure
can be, for instance,  the Shannon entropy, but also other measures could make sense.

In \cite{dominikK} the probabilistic setting is generalized to the case where every observation is formalized by a
binary  string $x_j$ (without any statistical population).  The information content of an observation is then measured using
Kolmogorov complexity (also ``algorithmic information'') which gives rise to an algorithmic version of (conditional) mutual information.
The corresponding functional model is given by  a Turing machine that computes the string $x_j$ from its parent strings $pa_j$ and a noise $n_j$.\\
The algorithmic information theory based approach generalizes the statistical framework since the average algorithmic information content per instance of a sequence of i.i.d. observations converges  to the Shannon  entropy, but on the other hand observations need not be generated by i.i.d. sampling.

Unfortunately, Kolmogorov complexity is uncomputable and  practical causal inference schemes must deal with other measures of information. In Section~\ref{SecGIM} we define general information measures and show that they induce independence relations that satisfy the semi-graphoid axioms (Section~\ref{secGraphoid}). Then, in Section~\ref{secMarkov}, we phrase the causal Markov condition within our general setting and  explore under which conditions 
it is a reasonable postulate. To this end, we formulate an information theoretic version of functional models observing that their decisive feature is that the joint information of a node, its parents and its noise is the same as the joint information of its parents and noise alone. We demonstrate with examples how these functional models restrict the set of allowed causal mechanisms to a certain class (Section~\ref{secEx}). We emphasize that the choice of the information measure determines this class and is therefore the essential prior decision 
(which certainly requires domain knowledge). Thus, when applying our theory to real data, one first has to think about the causal mechanisms to be explored and then design  an information measure that is sufficiently ``powerful'' to detect the generated dependences.\\
Section~\ref{secFaith} discusses a modification for known independence based causal inference that  is necessary for those information measures for which conditioning can only decrease dependences. Section~\ref{secCompress} describes one of the most important intended applications of our theory, namely information measures based on compression schemes (e.g. Lempel-Ziv). Applications of these measures using the PC algorithm for causal inference to segments of English text demonstrate the strength of causal reasoning that goes beyond already known applications of compression for the purpose of (hierarchical) clustering.

\section{General information measures}

\label{SecGIM}

In this section we define information from an axiomatic point 
of view and prove properties that will be useful in the derivation of the causal Markov condition. We start by rephrasing  the usual concept of measuring statistical dependences. Let ${\cX}$ be a set of discrete-valued random variables and
$\Omega:=2^{\cX} $ be  the set of subsets. For each  $A\in  \Omega$ let
$H(A)$ denote the joint Shannon entropy of the variables in $A$. For three disjoint 
sets $A,B,C$ the conditional mutual information between $A$ and  $B$ given $C$
then reads
\begin{equation}\label{condMu}
I(A:B|C):=H(A \cup C)+H(B\cup C) - H(A \cup B \cup C)-H(C)\,. 
\end{equation}
The set of subsets constitutes a lattice $(\Omega,\vee,\wedge)$ with respect to the operations of union and intersection and $H$ can be seen as a function on this lattice\footnote{Also the information function that are presented in this paper can all be rephrased as functions on the lattice of subsets it is nevertheless notationally convenient to formulate the theory with respect to general lattices.}. We observe that the non-negativity of (\ref{condMu}) can be guaranteed if
\[
H(D)+H(E)\geq H(D \vee E)+H(D\wedge E)\,, 
\]
for two sets $D,E\in  \Omega$. This {\it submodularity}  condition
is  known to be true for Shannon entropy  \cite{coverThomas}. Motivated by these remarks, we now  introduce an abstract information 
measure defined on the elements of a general
lattice. Throughout this paper let $(\Omega,\wedge,\vee)$ be a finite lattice and denote by $0$ the meet of all of its elements.

\begin{Def}[information measure]$ $\\We say
$
R: \Omega \rightarrow \mathbb{R}_+
$
is an \emph{information measure} if it satisfies the following axioms:
\vspace{0.1cm}

\noindent
(1) normalization:\quad
$
R(0)=0\,,
$
\vspace{0.1cm}

\noindent
(2)
monotonicity:\quad\;
$s \leq  t$ \quad implies \quad   $R(s) \leq R(t)$\quad for all $s,t\in \Omega$\,,
\vspace{0.1cm}

\noindent
(3) submodularity:\quad
$
R(s)+R(t)\geq R(s\vee t)+R(s \wedge t)$  for all $s,t\in \Omega$\,.
\end{Def}
Note that submodular functions have been considered in different contexts for example in \cite{lovasz} and \cite{matus94}.\\
Based on $R$ we define a conditional version for all $s,t\in\Omega$ by
\[
R(s|t) := R(s\vee t)-R(t).
\]
For ease of notation we write $R(s,t)$ instead of $R(s\vee t)$. In analogy to (\ref{condMu}), $R$ gives rise to the following measure of independence.

\begin{Def}[conditional mutual information]${}$ \label{def:cmi}
For $s,t,u\in\Omega$ the conditional mutual information of $s$ and $t$ given $u$ is defined by
\begin{eqnarray*}
I(s:t|u) &:=& R(s,u)+R(t,u)-R(s,t,u)-R(u).
\end{eqnarray*}
We say $s$ and $t$ are independent given $u$ or equivalently $\;\;s \independent t\,|u\;\;$ if $ I(s:t|u)=0$. 
\end{Def}
The following Lemmas generalize usual information theory.

\begin{Lem}[non-negativity of mutual information and conditioning]${}$
For $s,t,u\in\Omega$ we have
\[
\hbox{(a)} \quad   I(s:t|u) \geq 0   \quad \quad  \hbox{ and }\quad \quad  \hbox{(b)} \quad  0 \leq R(s|t,u) \leq R(s|t). \hspace{7cm}\,
\]
\end{Lem}
Proof: $(a)$ By definition,
$
I(s:t|u)\geq 0$  is equivalent to  $R(s,u)+R(t,u) \geq R(s,t,u) + R(u)$.
Defining $a=s\vee u$ and $b=t\vee u$ and using associativity of $\vee$ we have $a\vee b=s\vee t\vee u$. Further, using Lemma 4 in Ch.1 from \cite{birkhoff}, in any lattice
$$
a \wedge b = (s\vee u) \wedge (t\vee u) \geq  u \vee (s\wedge t) \geq u
$$
and hence by monotonicity of $R$: $R(a \wedge b)\geq R(u)$. Combining everything 
$$
R(s,u)+R(t,u) = R(a)+R(b) \geq  R(a\vee b) + R(a\wedge b) \geq R(s,t,u)+R(u),
$$
where the first inequality uses submodularity of $R$.\\
$(b)$ The first inequality follows from $(a)$ by $I(s:s|t,u)\geq 0$. The second inequality follows directly from $(a)$ and the definition of $I$.
$\Box$

\begin{Lem}[chain rule for mutual information]${}$\label{lemChain}
For $s,t,u,x\in\Omega$
\begin{equation}\label{chainRule}
I(s: t\vee u|x)=I(s:t|x)+I(s:u|t,x)
\end{equation}
\end{Lem}
Proof: This is directly seen by using the definition of conditional mutual information on both sides.$\Box$

\begin{Lem}[data processing inequality]${}$
Given $s,t,x \in \Omega$ it holds
$$
R(s|t)=0 \hspace{0.5cm} \Rightarrow \hspace{0.5cm} I(s:x|t) = 0 \hspace{0.5cm} \Rightarrow \hspace{0.5cm} I(s:x) \leq I(t:x).
$$
\end{Lem}
Proof: The first implication is clear. For the second we apply the chain rule for mutual information two times and obtain
\begin{eqnarray*}
I(s:x) &=& I(s,t:x)-I(t:x|s) = I(t:x) + I(s:x|t) - I(t:x|s) \leq I(t:x)\,,
\end{eqnarray*}
since the second summand is zero by assumption and conditional mutual information is non-negative.
$\Box$
\vspace{0.5cm}

\section{Submodular dependence measures and semi-graphoid axioms}

\label{secGraphoid}

The axiomatic approach to stochastic independence goes back to Dawid~\cite{dawidCI} who stated four axioms of conditional independence that are fulfilled for any kind of probability distribution. Later, any relation $I$ on triplets that satisfies the same axioms has been named semi-graphoid in \cite{pearlCausality}. In the following we show that the function $I$ constructed from $R$ in the last section satisfies these axioms.

\begin{Lem}[$I$ satisfies semi-graphoid axioms]\label{lemGraphoid}
The function $I$ defined in the last section satisfies the semi-graphoid axioms, namely for $x,y,w,z\in \Omega$
\[
\begin{array}{lrcrl}
   (1) &  I(x:y|z) = 0 \quad \,& \quad \quad\quad  \Rightarrow \quad \quad \quad  & I(y:x|z)=0 \,\,\,\, & \hbox{(symmetry)} \\
 (2) & I(x:y,w|z)=0 \quad\,  & \quad\quad\quad \Rightarrow \quad\quad\quad & \left\{\begin{array}{r} I(x:y|z)=0\, \\ I(x:w|z)=0 \end{array}\right. & \hbox{(decomposition)}\\
(3) & I(x:y,w|z)=0 \quad\, & \quad \quad\quad\Rightarrow \quad \quad\quad & I(x:y|z,w)=0 \,\,\,\,& \hbox{(weak union)}\\
(4) & \left.\begin{array}{r} I(x:w|z,y)=0 \\ I(x:y|z)=0
    \end{array}\right\} & \quad\quad\quad \Rightarrow\quad\quad\quad  & I(x:w,y|z)=0 \,\,\,\,& \hbox{(contraction)}
\end{array}
\]

\end{Lem}
Proof: Symmetry is clear and the remaining implications follow directly from the chain rule and non-negativity.
$\Box$
\vspace{0.4cm}

On the contrary, if we are given a function $I: \Omega\times\Omega\times\Omega \rightarrow \mathbb{R}_+$, what axioms do we need to define a submodular information measure $R$ from $I$? It turns  out that the chain rule in eq.~(\ref{chainRule})
together with non-negativity $I(a:b|c)\geq 0$ and symmetry $I(a:b|c)=I(b:a|c)$
already implies  that
$R(a) := I(a:a|0)$ is an information measure and $I$ coincides
with the dependence measure introduced  in Definition~\ref{def:cmi}.
We omit  the  proof due to space  constraints.

Thus we characterized the type of dependence measures that we are able to incorporate into our framework. To show that the chain rule is actually a strong restriction we close the section with an example of independence of orthogonal linear subspaces.\\

\noindent\textbf{Example: (orthogonal subspaces, qualitative version e.g. in \cite{lauritzen96})}\\
Linear subspaces of some finite vector space form a lattice, where the union of two subspaces is the subspace generated by the set-theoretic union. For two such subspaces $a$ and $b$ write $\pi_b (a)$ for the orthogonal projection of $a$ onto $b$. We define $I(a:b) = \dim \pi_b (a)$, hence orthogonal subspaces are considered as independent. Given a third subspace $c$ the conditional version reads 
$$
I(a:b|c) = \dim \pi_{b_{|c^{\bot}}} (a_{|c^{\bot}}),
$$
where $a_{|c^{\bot}}$ and $b_{|c^{\bot}}$ denote the orthogonal projections of $a$ and $b$ to the orthogonal complement $c^{\bot}$ of $c$. $I$ is symmetric and non-negative, but it does not satisfy the chain rule. This is because projecting a subspace first to $c^{\bot}$ and then to $b^{\bot}$ is different from projecting it to $(b\vee c) ^{\bot}$. Hence we can not find a submodular function $R$ underlying our dependence measure. Nevertheless $I$ satisfies the semi-graphoid axioms and can thus be considered as a measure of independence. 

\section{Causal Markov condition for general information measures}\label{secMarkov}

In this section we define three versions of the 
causal Markov  condition with respect to
a  general submodular information measure and show that they are equivalent
(similar to the statistical framework).
Then we discuss under which conditions we expect it to be a reasonable postulate
that links observations with causality. 
Assume we are given observations $x_1,\ldots,x_k$ that are connected by a DAG.  It is no restriction to consider the observations as elements of a lattice, e.g. the lattice of their subsets.

\begin{Def}[causal Markov condition (CMC), local version]  Let $G$ be a $DAG$ that describes the causal relations among observations $x_1,\ldots,x_k$. Then 
the observations are said to fulfill the causal Markov condition
with  respect to
the dependence measure $I$ if
\[
I(nd_j:x_j|pa_j)=0\hspace{0.5cm} \text{ for all }\;1\leq j\leq k,
\]
where $pa_j$ denotes the join of the parents of $x_j$ and $nd_j$ the join of its non-descendants (excluding the parents).
\end{Def}
The intuitive  meaning of the postulate is that conditioning on the direct causes of an observation screens off its dependences from all its non-effects. The following theorem generalizes results in \cite{lauritzen96} for statistical independences  and \cite{dominikK} for algorithmic independences. In particular it states that if the causal Markov condition holds with respect to a graph $G$, then independence relations implied by the CMC can be obtained through the convenient graph-theoretical criterion of $d$-separation~ (\cite{pearlCausality}, \cite{spirtes}).

\begin{Theorem}[Equivalence of Markov conditions and information decomposition]\label{thmMarkovEquiv}$ $\\
Let the nodes $x_1,\ldots,x_k$ of a DAG $G$ be  
elements of some lattice $\Omega$ and  $R$ be 
an information measure on $\Omega$. Then the following three properties are equivalent
\begin{enumerate}
\item[(1)] $x_1,\ldots,x_k$ fulfill the (local) causal Markov condition.
\item[(2)] For every ancestral set\footnote{A set $A$ of nodes of a DAG G is called ancestral, if for every $v \in A$ the parents of $v$ are in $A$ too.} $A\subseteq \{x_1,\ldots,x_k\}$, $R$ decomposes according to $G$: 
\[
R(A) = \sum_{x_i\in A} R(x_i|pa_i).
\]
\item[(3)] The global Markov condition holds, i.e., if three sets of nodes $A,B,C$ are d-separated in $G$, then
$$
\left(\bigvee_{a\in A}a\right) \quad  \independent  \quad \left(\bigvee_{b\in B}b \right)  \quad \big| \quad \left(\bigvee_{c\in C}c\right)\,.
$$
\end{enumerate}
\end{Theorem}

The proof is provided in Appendix \ref{apThm1}. The second condition shows that the joint information of observations can  be  recursively  computed according to the causal structure. The third condition describes explicitly which sets of independences are implications  of the causal Markov condition.\\

\begin{figure}
  \begin{center}
    \includegraphics{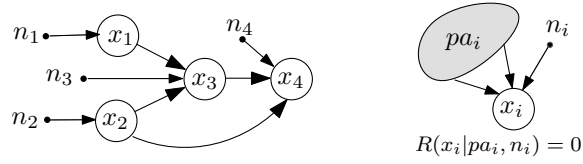}
  \end{center}
  \caption{  \label{figFunc}\small On the left a causal model of four observations $x_1,\ldots,x_4$ is shown together with the 'noise' for each node. In Lemma $\ref{lemExt}$ it is shown that the causal Markov condition on this extended graph implies the CMC for $x_1,\ldots,x_4$. On the right hand side the functional model assumption is illustrated: The generation of $x_i$ from its parents $pa_i$ and the 'noise' does not produce additional information.}

\end{figure}

Our next Theorem will  show that the CMC follows from a general notion of a functional model. At its basis is the following Lemma describing that the CMC on a given set of observations can be derived from the causal Markov condition with respect to an extended causal graph (see Figure \ref{figFunc}).\\
\begin{Lem}[causal Markov condition from extended graph]$ $\label{lemExt}
Let the nodes $x_1,\ldots,x_k$ of a DAG $G$
be elements of a lattice $\Omega$ with an independence relation $I$ that is monotone and satisfies the chain rule. 
If there exist additional elements $n_1,\ldots,n_k \in \Omega$ such that for all $j$
\begin{equation}\label{ass1}
I(x_j\,:\,nd_j,n_{-j}\,|\,pa_j, n_j)=0\,,  \quad  \quad  \text{ where }\;\; n_{-j}= \bigvee_{i\neq j} n_i, 
\end{equation}
and the $n_j$ are jointly independent in the sense that
\begin{equation}\label{ass2}
I(n_j : n_{-j}) = 0\,,
\end{equation}
then  the $x_1,\dots,x_k$ fulfill the causal Markov condition with respect to $G$.
\end{Lem}
Proof: Based on $G$ we construct a new graph $G'$ with node set $\{n_1,\dots,n_k\}\cup \{x_1,\dots,x_k\}$ and an additional edge $n_j \rightarrow x_j$ for every $j, (1\leq j\leq k)$. We first show that the causal Markov condition holds for the nodes of $G'$: By construction, the join of non-descendants $nd'_j$ of $x_j$ with respect to $G'$ is equal to $n_{-j}\vee nd_j$. Since the join of the parents $pa'_j$ of $x_j$ in $G'$ are $pa_j\vee n_j$, assumption (\ref{ass1}) just states $I(x_j : nd_{j}' | pa_j')=0$ which is the local CMC with respect to $x_j$. To see that CMC also holds for $n_j$, observe that the non-descendants of $n_j$ are equal to the non-descendants of $x_j$ in $G'$ and since $n_j$ does not have any parents, we have to show
\begin{equation}\label{eqMarkN}
I(n_j : nd'_j) = 0.
\end{equation}
Using $nd'_j = n_{-j}\vee nd_j$ together with the chain rule for mutual information we get
$$
I(n_j : nd_j,n_{-j})  = I(n_j:n_{-j})+I(n_j:nd_j|n_{-j}) = I(n_j:nd_j|n_{-j}),
$$
where the last equality follows from (\ref{ass2}). Let $ND_j=\{x_{j_1},\ldots, x_{j_{k_j}}\}$ be the set of non-descendants of $x_j$ in $G$. Note that $ND_j$ is ancestral, that is if $x \in ND_j$, then so are the ancestors of $x$. We introduce a topological order on $ND_j$, such that if there is an edge $x_{j_a}\rightarrow x_{j_b}$ in $G$, then $x_{j_a} < x_{j_b}$. Using the chain rule for mutual information iteratively we get
$$
I(n_j:nd_j|n_{-j}) = \sum_{a=1}^{k_j} I\big(n_j:x_{j_a} | x^{(<)}_{j_{a}}, n_{-j}\big),
$$
where $x^{(<)}_{j_{a}}$ denotes the join of elements of $ND_j$ smaller than $x_{j_{a}}$. By choice of our ordering the mutual information of $n_j$ and $x_{j_a}$ is conditioned at least on its parents and we can write $x^{(<a)}_{j_{a}} = pa_{j_a} \vee pa^c_{j_a}$, where $pa^c_{j_a}$ is the join of elements smaller than $x_{j_a}$ in $ND_j$ that are not its parents. Therefore, again by the chain rule, each summand on the right hand side can be bounded from above by writing
\begin{eqnarray*}
I\big(n_j:x_{j_a} | x^{(<a)}_{j_{a}}, n_{-j}\big) &\leq & I\big(n_{-j_a},pa^c_{j_a}:x_{j_a} | pa_{j_a}, n_{j_a}\big) \\
&\leq & I\big(n_{-j_a},nd_{j_a}:x_{j_a} | pa_{j_a}, n_{j_a}\big) = 0,
\end{eqnarray*}
where the second inequality is true because by construction $pa^c_{j_a}$ is the join of non-descents of $x_{j_a}$. The right hand side vanishes because of assumption (\ref{ass1}). This proves (\ref{eqMarkN}) and therefore the causal Markov condition with respect to $G'$.\\
By Theorem \ref{thmMarkovEquiv}, $d$-separation on $G'$ implies independence. Due to the special structure of $G'$ one can check that $d$-separation in $G$ implies $d$-separation in the extended graph $G'$.  Again by Theorem \ref{thmMarkovEquiv}, $d$-separation implies the causal Markov condition for $G$, which proves the lemma. $\Box$\\

Now we formalize the intuition that in a generalized functional model a node only contains information that is already  contained 
in the direct causes and the noise together (see Figure \ref{figFunc}):

\begin{Def}[functional model]
Let $G$ be a DAG with nodes $x_1,\dots,x_k$ in the lattice $\Omega$. 
If there exists an  additional node $n_j \in \Omega$ for each $x_j$, such that the 
$n_j$ are jointly independent and
\begin{equation}\label{eqFuncModel}
R(x_j,pa_j,n_j) = R(pa_j,n_j) \hspace{0.5cm} \text{ for all } j, (1\leq j\leq k)
\end{equation}
then $G$ together with $n_1,\dots,n_k$ 
is called a \emph{functional model} of the  $x_1,\dots,x_k$.
\end{Def}
If we restrict our attention to causal mechanism of the above form, the CMC is justified:

\begin{Theorem}[functional model implies CMC]\,$ $\label{thmMark}
If there exists a functional model for the nodes $x_1,\dots,x_k$ of a DAG $G$ 
then they fulfill the causal Markov condition with respect to $G$.
\end{Theorem}
Proof: In the functional model with noise nodes $n_i$ it holds $R(x_j,pa_j,n_j) = R(pa_j,n_j)$ for all $j$. This implies $I(nd_j:x_j|pa_j\vee n_j)=0$. Since the $n_j$ in a functional model are assumed to be jointly independent, Lemma \ref{lemExt} can be applied and proves the theorem. $\Box$\\

The following 
section describes examples of causal mechanisms that can be seen as functional models with respect to various information measures.

\section{Examples of information measures and their functional models}\label{secEx}

Let $x_1,\ldots,x_k$ be a finite set of observations which are in a canonical way elements of the lattice of subsets $(2^S,\cup,\cap)$.  Let the causal structure be a DAG with $x_1,\ldots,x_k$ as nodes.

\subsection{Shannon entropy of random variables}

Let the $x_i$ be discrete random variables with joint probability mass function $p(x_1,\ldots,x_k)$. For a subset $A\subseteq \{x_1,\ldots,x_k\}$ denote by $x_A:=\times_{x_i\in A}x_i$ the random variable with distribution $p_A:=p((x_i)_{x_i\in A})$. The Shannon entropy for the subset $A$ is defined as $H(A) := -\mathbb{E}_{p}\log p_A$. Monotony as well as  submodularity are well-known properties \cite{coverThomas}. The corresponding notion of independence is the familiar (conditional) stochastic independence, its information-theoretic quantification $I$ being mutual information. 
Then $H(x_i,pa_i,n_i)=H(pa_i,n_i)$ is equivalent to the existence of some function $f_i$ with
\[
x_i=f_i(pa_i,n_i)\,.
\]
This restricts the set  of mechanisms to those which were deterministic if one could take all  latent factors into account. Note that continuous Shannon entropy is  not monotone
under restriction to subsets.
Nevertheless, in this case the chain rule and non-negativity is true and therefore the CMC can be motivated by independences with respect to an extended causal model (Lemma \ref{lemExt} of the previous section).

\subsection{Kolmogorov complexity  of binary strings}

Let the $x_i$ be binary strings and the information measure be the Kolmogorov complexity as information measure. More explicitly, for a subset of strings $A\subseteq S$ denote by $x_A$ a concatenation of the strings  in a prefix free manner (which guarantees that the concatenation can be uniquely decoded into its components). 
The Kolmogorov complexity $K(x_A)$ is then defined as the length of the shortest program that generates the concatenated string $x_A$ on a universal prefix-free Turing machine. It is submodular up to a logarithmic constant \cite{ham00}. 
For  two strings $s,t$ the conditional Kolmogorov complexity $K(s|t)$ of $s$,   given $t$
is defined as the length  of the shortest program that computes $s$ from the  input $t$.
It must  be distinguished from $K(s|t^*)$, the length of the shortest program that
computes $s$ from the shortest compression of $t$.
Note  that defining $R(s):=K(s)$ implies that  the conditional information
reads $R(s|t)=K(s|t^*)$ due   to 
\[
K(s,t)\stackrel{+}{=} K(t)+K(s|t^*)\,,
\]
see \cite{algStat}.
Then
\[
K(x_i,pa_i,n_i)\stackrel{+}{=}K(pa_i,n_i)\quad \hbox{  is equivalent to  } \quad
K(x_i|(pa_i,n_i)^*)\stackrel{+}{=}0 \,,
\]
which, in turn, is equivalent to
the existence of a program of length  $O(1)$ that computes $X_i$ from the shortest compression of
$(pa_i,n_i)$. Here we have considered the number $k$ of nodes  as  a constant, which 
ensures that the order of the strings  does  not matter.
Such an ``algorithmic model of causality''\cite{dominikK} 
restricts causal influences to {\it computable} ones.
Uncomputable mechanisms can easily be defined (halting problem).
However, in the spirit of the Church-Turing thesis, we will  assume  that they don't exist in nature
and conjecture that the algorithmic model of causality is the most general model of a  
causal mechanism as  long as we restrict the  attention to the non-quantum world
(where the  model would probably be replaced with a quantum  Turing machine).

\subsection{Period length of time series}

\label{sub:per}

We start with the following abstract example to illustrate that the definition of an information measure is more natural when the observations are taken to be part of a lattice different from the lattice of subsets. Let every observation be a natural number $x_i \in \N$  and consider them elements of the lattice of natural numbers where $\vee$ denotes the least common multiple and $\wedge$ the greatest common divisor, hence for $S\subseteq \{x_1,\ldots,x_k\}$
\[
x_S:=\vee_{x_i\in S} x_i :=lcm((x_i)_{x_i\in S}) \hspace{1cm} \,.
\]  
We define an information measure by
\[
R(x_S):=\log x_S\,.
\]
Non-negativity and monotonicity of $R$ are clear and submodularity even holds with equality: For $a,b\in\mathbb{N}$ 
\begin{eqnarray*}
R(a\vee b)+R(a \wedge b) &=& \log lcm(a,b)  + \log gcd(a,b)
= \log \frac{ab}{gcd(a,b)}  + \log gcd(a,b)\\
&=& R(a)+R(b).
\end{eqnarray*}
The corresponding conditional dependence measure reads
$$
I(a:b|c) =  R\big({gcd(a,b)}/{gcd(a,b,c)}\big) = \log gcd(a,b) - \log gcd(a,b,c),
$$
so $a$ and $b$ are independent given $c$ if $c$ contains all prime factors that are shared by $a$ and $b$ (with at least the same multiplicity).\\
We define a functional model where every node $x_i$ contains only
prime factors that are already contained in its parents and its noise node (with at least the same multiplicity) and the noise terms are assumed to be relatively prime.

Such a lattice of observations can occur in real-life if  $x_i$ denotes  the 
period length of a periodic time series over $\Z$. Then the period length 
of the joint time series defined by a set of nodes is obviously the least common multiple. If every time series at node $i$ 
is a function $F_i$ of its parents and noise node
(each being a time series) and $F_i$ is {\it time-covariant}, $x_i$ divides their period lengths.

Assuming  that the period lengths of the noise time series are  relatively prime
is indeed a strong restriction, 
but if we assume
that the  periods are large numbers and interpret independence in the approximate sense
\[
\log lcm(\{x_i\}) \approx \sum_i  \log x_i\,,
\]
we obtain the condition that their periods have no {\it large} factors in  common.
This seems to be a reasonable assumption if the noise time series have  no common  cause.

One can easily think of generalizations where every observation $x_i$  
is characterized by a symmetry group and the join of nodes by the group intersection
describing the joint symmetry. One may then define functional models where every node
inherits all those symmetries that are shared by all its parents and the noise node.

\subsection{Size of vocabulary in a text}

\label{sub:voc}

Let every observation $x_i$ be a text and for every collection of texts $S \subseteq \{x_1,\ldots,x_k\}$ let $R(S)$ be the number of different meaningful words in $S$.
Here, meaningful means that we ignore words like articles and prepositions.
To see that $R$ is submodular we observe that it is just the number of elements of a set.

We can use $R$ to explore which author has copied parts of the texts written by other authors: Let every $x_i$ be written by another author and a causal
arrow from $x_i$ to $x_j$ means that the author of $x_i$ was influenced by $x_i$ when writing $x_j$.

The noise $n_i$ can be interpreted as the set of words the author usually uses and 
the condition $R(x_i,pa_i,n_i)=R(pa_i,n_i)$ then means that he/she combines only words from the texts he/she has seen with the own vocabulary.

To conclude this section we want to emphasize that the above example refers to a dependence measure that is non-increasing under conditioning, that is for collections $S,T,U$ and $V$ of texts $I(S:T|U) \geq I(S:T|V)$ whenever $U\subseteq V$. This is because $I(S:T|U)$ is equal to the number of meaningful words contained in $S$ and $T$, but not in $U$. In general, the above information measure can be viewed as rank or height function (cardinality) on the lattice of sets of meaningful words and it can be shown that dependence measures originating form information functions that are rank functions on distributive lattices are always non-increasing under conditioning \footnote{Lattices with a rank function that is submodular are generally called semimodular lattices. Abstract independence measures on semimodular lattices have also been discussed in a different context by Cuzzolin~\cite{cuzzolin08}.}. We will elaborate on this point in the next section because it imposes special challenges for causal inference.

\section{Faithfulness for monotone dependence measures}\label{secFaith}

Apart from the CMC, the essential postulate of independence based causal inference is usually {\it causal faithfulness}. It states that all observed independence relations are structural, that is, they are induced by the true causal DAG through $d$-separation.  This postulate
allows the identification of causal DAGs up to ``Markov equivalence classes''
imposing the same independences. 

Faithfulness has already been defined for abstract conditional independence statements and we start by rephrasing the definition following (\cite{spirtes}, p.81).
\begin{Def}[faithfulness]\label{defFaith}
A DAG $G$  is said to represent a list of conditional independence relations $\mathcal{L}$ on a set of observations $X$ \emph{faithfully}, if $\mathcal{L}$ consists \emph{exactly} of the independence relations implied by $G$ through $d$-separation.
Further, a set of observations $X$ is said to be faithful (w.r.t. a given dependence measure), if there exists a causal model that represents $X$ faithfully.
\end{Def}
The above definition of faithfulness makes sense for the probabilistic and algorithmic notions of dependence, but there is a problem with respect to dependence measures on which conditioning can only decrease information. As 
mentioned above,
rank functions of distributive lattices lead to this kind of dependence measures, that we will call \emph{monotone} in the following. To see the problem, consider for three observations $a,b,c$ a causal model $G$ of the form $a\rightarrow b \leftarrow c$. By $d$-separation, $a$ is independent of $c$ and for a monotone dependence measure this implies $a \independent c|b$, which is not an independence induced by $d$-separation. Hence, $G$ does not faithfully represent the objects and one can easily check that a faithful representation does not exist (e.g. using the theorem below). However, we can modify  faithfulness
such that it also accounts for those independences that  follow from monotony
under conditioning: 
\begin{Def}[monotone faithfulness]
A DAG  $G$ is said to represent a list $\mathcal{L}$ of conditional independences of observations $X$ monotonely faithful, if the following condition is true for all disjoint subsets $S,T,U\subseteq X$ whose join is denoted by $s,t$ and $u$: 
Whenever $s \independent t |u$ is in $\mathcal{L}$ and $u$ is \emph{minimal} among all the sets that render $s$  and $t$ independent, then $s$  and $t$ are $d$-separated by $u$ in $G$. Further, a set of observations $X$ is said to be monotonely faithful (w.r.t. a given dependence measure), if there exists a causal model that represents $X$ monotonely faithful.
\end{Def}
Note that, trivially, every faithful representation is a monotonely faithful representation, hence faithful observations are monotonely faithful observations. Faithful representations have already been characterized (Theorem 3.4 in \cite{spirtes}) and we prove an equivalent characterization that holds simultaneously for monotonely faithful and for faithful observations.

\begin{Theorem}[characterization of monotonely faithful representations]\label{thmFaith}
A set of (monotonely) faithful observations $X$ is represented (monotonely) faithfully by a DAG $G$ if and only if
\begin{enumerate}
\item[(1)] two observations $a$ and $b$ are adjacent in $G$ if and only if they can not be made independent by conditioning on any join of observations in $X\backslash\{a,b\}$.
\item[(2)]  for three observations $a,b,c$, such that $a$ is adjacent to $b$, $b$ is adjacent to $c$ and $a$ is not adjacent to $c$, it holds that $a \rightarrow b \leftarrow c$ in $G$ if and only if there exists a set $U\subseteq X\backslash \{a,b,c\}$ such that $a$ is independent of $c$ given the join of the observations in $U$.
\end{enumerate}
\end{Theorem}
The proof is given in the appendix. 
The theorem implies in particular, that every monotonely faithful representation of faithful objects is already a faithful representation.\\
The PC algorithm \cite{spirtesPC,spirtes} for causal inference takes a set of conditional independences on faithful objects and returns the equivalence class of faithful representations. Since the above theorem is used to prove the correctness of the algorithm in the  faithful case, we conclude that the algorithm correctly returns monotonely faithful representations given monotonely faithful observations. We apply the PC-algorithm with respect to compression based information functions in the following section. Also they are not monotone in a strict theoretical sense, empirical observations indicate that it is unlikely for the mutual information to increase. 

\section{Compression based information}\label{secCompress}

In this section we demonstrate that our framework enables us to do causal inference on single \emph{objects} (coded as binary strings) without relying on the uncomputable measure of Kolmogorov complexity. To this end, instead of defining complexity with respect to a universal Turing machine we explicitly limit ourselves to specific production processes of strings. The underlying measure of information is motivated by universal compression algorithms like LZ77 \cite{lz77} and grammar based compression \cite{yang00} that detect repeated occurrences of identical substrings within a given input string and encode them more efficiently. The choice of a compression scheme can be seen as a prior analogously to the choice of a universal Turing machine in the case of algorithmic information. The measures considered in this section quantifiy the information of an observation (string) in terms of the diversity of its substrings and entail the following assumption on causal processes: A mechanism that produces a string  $y$ from a string $x$ is considered as simple, if it constructs $y$ by concatenating a small number of substrings from $x$ (see Lemma \ref{lemLZfun} below for a formal statement). Further, the amount of dependence of observations is approximately given by the number of substrings that they share.

We are going to describe two specific measures of information that are closely related to the total length of the compressed string, but have better formal properties than the latter. This way our conclusions will be independent of the actual implementation of the compression scheme and proving theoretical results gets easier.\\
In the last part of this section we describe experiments on real data in which the PC algorithm is applied to infer the causal structure using either of the two introduced measures of information.\\
Note that \emph{distance metrics} based on compression length have already been used to cluster various kinds of data (see \cite{clustering05} for computable distance metrics motivated by algorithmic mutual information or \cite{hanus07} for an application to molecular biology). These metrics can be used to reconstruct trees (hierarchical clustering) but if two nodes are linked by more than one path a measure of \emph{conditional} mutual information is needed to reconstruct the data-generation process. To the best of our knowledge, compression based methods have not been used before to infer non-tree-like DAGs.

\subsection{Lempel-Ziv information (LZ-information)}

LZ-information has been introduced as a complexity measure for strings in \cite{lz76}. It has been applied to quantify the complexity of time series in biomedical signal analysis \cite{aboy06} and  distance measures based on versions of LZ-information have been used to analyze neural spike train data \cite{lzNeuro} and to reconstruct phylogenetic trees \cite{zheng09}. We start by defining
\begin{Def}[production and reproduction from prefix]
Let $s=xy$ be a string. We say $s$ is \emph{reproducible} from its prefix $x$ and write $x\rightarrow s$ if $y$ is a substring of $x\overline{y}$, where $\overline{y}$ is equal to $y$ without its last symbol. We say $s$ is \emph{producible} from $x$ and write $x \Rightarrow s$ if $x\rightarrow \overline{s}$, where $\overline{s}$ is equal to $s$ without its last symbol.
\end{Def}
Contrary to reproducibility, producibility allows for the generation of new substrings, for if $x\Rightarrow s$, the last symbol of $s$ can be arbitrary.\\

\vspace{0.2cm}
	\begin{tabular}{l p{0.75\textwidth}}
    {\includegraphics{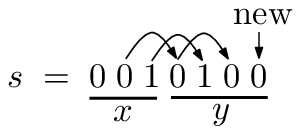}} & \vspace{-1.2cm}Example: For a given string $s=xy$ let $\overline{s}$ be the string without its last symbol. The figure on the left shows that $\overline{s}$ is producible from its prefix $x$ by copying the second symbol of $x$ to the first of $y$ and so on. The string $s$ itself not producible from $x$, but reproducible.\\
	\end{tabular}
\vspace{0.2cm}\\
Informally, LZ-information counts the minimal number of times during the process of parsing the input string from left to right, in which the string can not be reproduced from its prefix and a production step is needed. 
\begin{Def}[LZ-information, \cite{lz76}]\label{defLZ}
Let $s$ be a string of length $n$. Denote by $s_i$ the $i$-th symbol of $s$ and by $s(i,j)$ the substring $s_i s_{i+1}\cdots s_j$. A \emph{production history} $H_s$ of $s$ is a partition of $s$ into substrings $s=s(h_0,h_1)s(h_1+1,h_2)\cdots s(h_k+1,h_{k+1})$ with $h_0=1$ and $h_{k+1}=n$, such that
$$
s(1,h_i) \Rightarrow s(1,h_{i+1}) \quad \text{ for all } i \in \{1,\ldots,k\}.
$$
A history $H_s$ is called \emph{exhaustive} if additionally
$$
s(1,h_i) \not\rightarrow s(1,h_{i+1}) \quad \text{ for all } i \in \{1,\ldots,k-1\}.
$$
The substrings $s(h_i+1,h_{i+1}),\;(0\leq i\leq k)$ will be called components of $H_s$ and the length $|H_s|$ of $H_s$ is defined as the number of its components.\\
The \emph{LZ-information} of $s$, denoted by $c(s)$, is defined as the length of its (unique) exhaustive history.
\end{Def}
In an exhaustive history, each $h_i$ is chosen maximal such that $s(1,h_i-1)$ is reproducible from its prefix $s(1,h_{i-1})$. As an example, for $s=000100101100110$ the exhaustive history partitions $s$ into
$$
s = (0)(001)(00101)(10011)(0),
$$
hence $c(s)=5$.\\
In the original paper of Ziv and Lempel~\cite{lz76} it was shown that $c$ is subadditive: for two strings $x$ and $y$ the information of the concatenated string $xy$ is at most the information of $x$ plus the information of $y$. This already suggests the non-negative unconditional dependency measure $i(x:y)=c(x)+c(y)-c(xy)$. As it turns out, submodularity holds up to a  negligible constant independent of the involved string lengths:
\begin{Lem}[submodularity of  LZ-information, asymmetric version]\label{lemLZ}
Let $x,y,z$ be finite strings over some alphabet $\mathcal{A}$. Further let $\alpha$ and $\beta$ be symbols not contained in $\mathcal{A}$ that will be used as separators. Then
\begin{equation}\label{ineqLZ}
i(x:y|z) := c(z\alpha x) + c(z\alpha y) - c(z\alpha x\beta y)- c(z) \geq -1.
\end{equation}
\end{Lem}
Proof: Let $E_{z\alpha}$ be the exhaustive history of $z\alpha$. The exhaustive history of  $z\alpha x$ is of the form $E_{z\alpha x}=[E_{z\alpha},E_{x|z}]$, where $E_{x|z}$ describes the partition of $x$ induced by  $E_{z\alpha x}$. This is because $\alpha$ is not part of the alphabet, hence the component in $E_{z\alpha x}$ containing $\alpha$ must be of the form $(t\alpha)$ for some substring $t$. Analogously $E_{z\alpha y} = [E_{z\alpha},E_{y|z}]$. It is not difficult to see that
$$
H_{z\alpha x\beta y} = [E_{z\alpha},E_{x|z},\beta,E_{y|z}].
$$
is a production history of $z\alpha x\beta y$. Theorem $1$ in \cite{lz76} states that a production history is at least as long as the exhaustive history, hence
$$
\big|[E_{z\alpha},E_{x|z},\beta,E_{y|z}]\big| \geq |E_{z\alpha x\beta y}| = c(z\alpha x\beta y),
$$
Further,  $c(z) \leq |E_{z\alpha}|$ and so (\ref{ineqLZ}) can be bounded from below by
\begin{eqnarray*}
c(z\alpha x) + c(z\alpha y) - c(z\alpha x\beta y)- c(z)  &\geq & \big|[E_{z\alpha},E_{x|z}]\big| + \big|[E_{z\alpha},E_{y|z}]\big| - \big|[E_{z\alpha},E_{x|z},\beta,E_{y|z}]\big| - |E_z|\\
&=& -1.
\end{eqnarray*}
$\Box$
\begin{Lem}[functional model for LZ-information, asymmetric version]\label{lemLZfun} Let $pa_i$ and $n_i$ be two strings over an alphabet $\mathcal{A}$ and construct a third string string $x_i$ by concatenating $k$ substrings of $pa_i$ and $n_i$. Then 
$$
c(pa_i\,\alpha\,n_i\,\beta\,x_i) \leq c(pa_i\,\alpha\,n_i\beta) + k,
$$
where $\alpha$ and $\beta$ are symbols not in $\mathcal{A}$ used as separators.
\end{Lem}
Proof: A production history of $pa_i\alpha n_i\beta x_i$ can be generated by concatenating the exhaustive history of $pa_i\alpha n_i \beta$ with the list of the at most $k$ substrings out of which $x_i$ is constructed. The length of this history is $c(pa_i\alpha n_i) + k + 1$ and bounds $c(pa_i\alpha n_i\beta x_i)$ from above by Theorem $1$ in \cite{lz76}. $\Box$\\

In particular, if $xy$ is producible from $x$, by appending $y$, the information is at most increased by one. Hence, if we restrict the mechanisms that generate a node to consist of a limited number of concatenations of substrings from its parents and the independent noise  (compared to the amounts of information involved) the causal Markov condition would follow if $c$ were an information function. This is not the case since $c$ is not defined on sets of strings (in particular it is not symmetric ($c(xy)\neq c(yx)$), therefore  we define the LZ-information of a set of strings to be the LZ-information of their concatenation with respect to a given order (e.g. lexicographic).
\begin{Def}[LZ-information, set version]
Let $\{x_1,\ldots,x_k\}$ be a set of strings over some alphabet $\mathcal{A}$. Choose $k$ distinct symbols $\alpha_1,\ldots,\alpha_k$ not contained in $\mathcal{A}$ that will be used as separators.\\
Let $X=\{x_{i_1},\ldots,x_{i_m}\}$ be a subset and assume $x_{i_1}\leq x_{i_2}\leq \ldots \leq x_{i_m}$ with respect to a given order on the set of strings over $\mathcal{A}$. We define the LZ-information of $X$ as
$$
LZ(X) = c\big(x_{i_{1}}\,\alpha_{i_{1}} \cdots x_{i_{m}}\,\alpha_{i_{m}}\big),
$$
where the argument of $c$ is understood as the concatenation of the strings.
\end{Def}
Because of the asymmetry of $c$, $LZ$ is not monotone and submodular in a strict sense. However, empirical observations suggest that for sufficiently large strings the violations of submodularity induced by the asymmetries like $c(x\alpha y)\neq c(y \alpha x)$ are negligible compared to the amounts of information.\\

\noindent\textbf{Hypothesis:}\quad For practical purposes $LZ(\cdot)$ is an information measure up to constants at most logarithmic in the string length. The associated independence measure $I$ is monotonely decreasing (through conditioning).\\

We close by mentioning that the calculation of the LZ-information is very inefficient for large strings since one has to search over all substrings of the part of the string already parsed. In our implementation we therefore considered only substrings of length limited by a constant (we chose $30$ for strings of English text, since it is unlikely that a substring of length $30$ is repeated exactly).

\subsection{Grammar based information}

In the grammar based approach to compression an input string $x$ is transformed into a context-free grammar that generates $x$. This grammar is then compressed for example using arithmetic codes. We discuss this approach because it has been successfully applied to compress RNA data (e.g. \cite{liu08}). Further the LZ-based compression discussed in the previous section can be rephrased into this framework. As there are many grammars that produce a given string, it is essential that the transformation of strings to grammars produces economic representations of $x$ (for an overview see \cite{lehman02}) We implemented the so called greedy grammar transform from Yang and Kieffer~\cite{yang00}. It constructs the grammar iteratively by parsing the input string $x$. Due to space restrictions we just give an example of a string and its generated grammar.\\
\textbf{Example: } The binary string $x = 1001110001000$ is transformed using the greedy grammar transform of \cite{yang00} to the grammar $G(x):$
\begin{eqnarray*}
s_0 &\rightarrow & s_111s_2s_2 \\ 
s_1 &\rightarrow & 100\\
s_2 &\rightarrow & s_10,
\end{eqnarray*}
where $s_0,s_1$ and $s_2$ are variables of the grammar and $x$ can be reconstructed by starting from $s_0$ and then iteratively substituting $s_i$ by the right hand side of each production rule above. The \emph{length of a grammar} $|G(x)|$ is defined as the sum of all symbols on the right of every production rule, so for the above example $|G(x)| = 10$. We view the length of the constructed grammar as information measure of the string that it produces and define analog to the LZ-information
\begin{Def}[grammar based information]
Let $\{x_1,\ldots,x_k\}$ be a set of strings over some alphabet $\mathcal{A}$. Choose $k$ distinct symbols $\alpha_1,\ldots,\alpha_k$ not contained in $\mathcal{A}$ that will be used as separators.\\
Let $X=\{x_{i_1},\ldots,x_{i_m}\}$ be a subset and assume $x_{i_1}\leq x_{i_2}\leq \ldots \leq x_{i_m}$ with respect to a given order on the set of strings over $\mathcal{A}$. We define the grammar based information of $X$ as
$$
GR(X) = \big|G\big(x_{i_{1}}\,\alpha_{i_{1}} \cdots x_{i_{m}}\,\alpha_{i_{m}}\big)\big|,
$$
where the input of the grammar construction $G$ is understood as the concatenation of the strings.
\end{Def}
By definition $GR$ is non-negative and due to the construction process of the grammar it is monotone. However, experiments show that submodularity is violated, but the amount of violation still allows to draw causal conclusions for sufficiently large strings.

\subsection{Experiments}
This section reports the results on causal inference using the introduced LZ-information and grammar based information measures. 

\subsubsection*{Experiment 1: Markov chains of English texts}

We start with a string of English text $s_0$ from which we construct further strings $s_1,\ldots,s_k$ as follows: To generate $s_{i+1}$  we translate  $s_i$ using an automatic translator from Google\footnote{accessible at http://translate.google.de/} to a randomly chosen European language. Then $s_{i+1}$ is defined as the string that we obtain when we translate $s_i$ back to English using the same translator. Since $s_{i+1}$ is \emph{determined} by $s_i$, the process can be modeled by a 'Markov' chain $s_0\rightarrow \cdots \rightarrow s_k$. We then apply the PC algorithm\footnote{Our implementation of the PC algorithm for causal inference was based on the BNT-Toolbox for Matlab written by Kevin Murphy and available at http://code.google.com/p/bnt/.} to infer the corresponding equivalence class of (monotonely) faithful causal models consisting of the DAGs:
$$
s_0 \leftarrow \cdots \leftarrow s_i \rightarrow \cdots \rightarrow s_k \quad\quad \text{for}\quad  0\leq i\leq k.
$$
In our experiments we chose several starting texts of  $1000$ to $5000$ symbols (e.g. news articles and the abstract of this paper) and generated three strings $(k=3)$ using the described procedure. In every string we  transformed all non-space characters to numbers $0,\ldots,8$ using a modulo operation on the ASCII value to reduce the alphabet size. Repeated spaces were deleted and the space character has been encoded separately by the number $9$ to ensure that words of the string remain separated.

\noindent
\textbf{Results: } Based on the two information measures, the PC algorithm returned the correct class of DAGs in every case. For LZ-information the chosen threshold used to determine independence did not even have to depend on the starting texts $s_0$. Grammar based information seems to be more sensitive to the string lengths involved and we had to choose a different threshold for every chosen text $s_0$. Further, we successfully tried the method on the chain of preliminary versions of the abstract of this paper.

Finally note that methods based on compression distance could also be applied to recover the correct equivalence class. The crucial difference to our approach consists in the fact that we did not have to assume that the underlying graph is a tree.

\subsubsection*{Experiment 2: Four-node networks}

We want to infer the equivalence classes of (monotonely) faithful causal models depicted in Figures (a) and (b) below. To this end we randomly choose segments of a large English text and then construct the strings corresponding to the nodes $a,b,c$ and $d$ in a way that ensures the resulting observation $\{a,b,c,d\}$ to be (monotonely) faithful. Explicitely, we choose segments $s_x$ and $s_{xy}$ for each node $x$ and for each edge between nodes $x$ and $y$ respectively. Further, for every ordered triple of nodes $(x,y,z)$ whose subgraph is not equal to $x\rightarrow y \leftarrow z$, we pick a segment $s_{xyz}$. This way we obtain the following segments with respect to the graph in Figure (a):
$$
s_a,s_b,s_c,s_d,s_{ab},s_{ac},s_{bd},s_{cd},s_{bac},s_{abd},s_{acd}
$$
and with respect to the graph in Figure (b) we get segments
$$
s_a,s_b,s_c,s_d,s_{ac},s_{ad},s_{bc},s_{bd},s_{cad},s_{cbd}.
$$
Finally, the string at a node is constructed as the concatenation of all segments that contain the name of the node in its index (the order is arbitrary), e.g. in the case of Figure (a)
$$
b=s_bs_{ab}s_{bd}s_{bac}s_{abd}.
$$

As text source we chose an English version of Anna Karenina by Lev Tolstoi \footnote{The text is available at http://www.gutenberg.org/etext/1399.}. We then transformed all non-space characters to numbers from $0,\ldots,8$ using a modulo operation on the ASCII value to reduce the size of the alphabet. Repeated spaces were deleted and the space character has been encoded separately by the number $9$ to ensure that words of the string remain separated. The resulting string consisted of a total of approximately two million symbols. Using the above construction, we generated $100$ observations $\{a,b,c,d\}$ with respect to each graph and applied the PC algorithm. The length $N$ of the randomly chosen segments was chosen uniformly between $100$ and $200$ in the first run and between $300$ and $500$ in the second run. The choice of the threshold to determine independence depended only on the information measure and on the two possible ranges of $N$, but not on the indivudual observations. Further, the graph of Figure (b) implies an unconditional independence of $a$ and $b$. Since two disjoint segments of English text can not be expected to be independent, we conditioned all  informations that we calculate on  background knowledge in terms of fixed segment of length $5000$.

\begin{center}
\begin{tabular}{c p{0.24\textwidth} c p{0.24\textwidth}}
\includegraphics{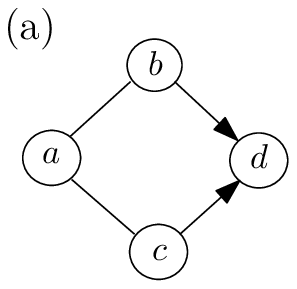} & \vspace{-3cm} 
\begin{tabular}{l r}
\multicolumn{2}{l}{Correct answers of PC:\vspace{0.1cm}}\\

\multicolumn{2}{l}{$N \in [100,200]$}\\
$LZ:$ & $98\%$\\  
$GR:$ & $53\%$\\  
$ $&$ $ \\
\multicolumn{2}{l}{$N \in [300,500]$}\\
$LZ:$ & $100\%$\\ 
$GR:$ & $56\%$ 
\end{tabular}
& \includegraphics{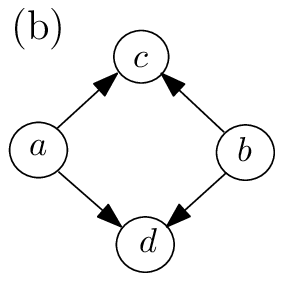} & \vspace{-3cm} 
\begin{tabular}{l r}
\multicolumn{2}{l}{Correct answers of PC:\vspace{0.1cm}}\\
\multicolumn{2}{l}{$N \in [100,200]$}\\
$LZ:$ & $95\%$\\ 
$GR:$ & $97\%$\\ 
$ $ & $ $ \\
\multicolumn{2}{l}{$N \in [300,500]$}\\
$LZ:$ & $100\%$\\ 
$GR:$ & $99\%$ 
\end{tabular}
\end{tabular}
\end{center}

\noindent
\textbf{Results: } Above, the percentages of correct results from the PC-algorithm are shown. Note that using LZ-information we were able to recover the correct equivalence class in almost all runs independently of the graph structure and segment length. Grammar based inference did not perform quite as well, but in the majority of cases in which it did not return the correct Markov equivalence class most of the independences still were detected correctly.

\section{Conclusions}

We have 
introduced conditional 
dependence measures that originate from submodular measures of information. 
We argued that these notions of conditional dependence 
(generalizing statistical dependence)
can be used to infer the causal structure among observations even if the latter
are not generated by i.i.d. sampling. 
To this end, we formulated a generalized causal Markov condition (with significant formal analogies to the statistical one)
 and proved that the condition is justified provided that the attention is restricted 
to a class of causal mechanisms that depends on the underlying measure of information. We demonstrated that existing compression schemes like Lempel-Ziv define interesting  notions of information and described the class of mechanisms that justify the causal Markov condition in this case.
Accordingly, we showed that the PC-algorithm successfully infers causal relations among texts when based notion of dependence induced by compression schemes.

\newpage
\appendix

\section{Proof of Theorem \ref{thmMarkovEquiv}}\label{apThm1}

$(1)\rightarrow (2)$. Let $A \subseteq \{x_1,\ldots,x_k\}$ be an ancestral set with respect to $G$ and denote by $G_A$ the subgraph of $G$ with nodes from $A$. Then the nodes of $A$ fulfill the causal Markov condition with respect to $G_A$: Denote by $ND_j$ the set of non-descendants of $x_j$ in $G$. Then we have for all $x_j\in A$
\begin{eqnarray*}
0= I(nd_j:x_j|pa_j) &\geq & I(nd_j^A:x_j|pa_j)= I(nd_j^A:x_j|pa^A_j),
\end{eqnarray*}
where $pa^A_j$ and $nd_j^A$ denote the join of the parents and non-descendants of $x_j$ in $G_A$.
The first equality follows from the causal Markov condition with respect to $G$, the last 
one uses $pa_j = pa^A_j$ (because $A$ is ancestral).

The remaining proof is by induction on the size $|A|$ of the ancestral set. For $|A|=1$ it is obvious. Assume the statement is true for sets with $|A|=k-1$ nodes and let $G$ be a graph with $k$ nodes on which $R$ fulfills the Markov condition. Without loss of generality assume that $x_k$ has no descendants. Hence by assumption of  the induction
\begin{equation*}
R(x_1,\ldots,x_{k-1}) = \sum_{i=1}^{k-1} R(x_i|pa_i).
\end{equation*}
By definition of conditioning we get 
\begin{equation}\label{eq1}
R(x_1,\ldots,x_k) = R(x_1,\ldots,x_{k-1}) + R(x_k | x_1,\ldots,x_{k-1}).
\end{equation}
Since $x_k$ was chosen to have no descendants it follows that $nd^A_k\vee pa^A_k = x_1 \vee \ldots\vee x_{k-1}$. Therefore
\[
R(x_k | x_1,\ldots,x_{k-1}) = R(x_k | nd^A_k\vee pa^A_k) = R(x_k | pa^A_k)\,,
\]
where the last equality follows from the independence of $x_k$ from its non-descendants given its parents, which is implied by the causal Markov condition with respect to $G_A$. Using this relation in equation (\ref{eq1}) proves $(2)$.\\
$(2)\rightarrow (1)$. We prove the causal Markov condition for every node $x_j$. Using the definition of conditional mutual information we get
\begin{eqnarray*}
I(nd_j:x_j|pa_j) &=& R(nd_j,pa_j)+R(x_j|pa_j)-R(nd_j,pa_j,x_j).
\end{eqnarray*}
Denote by $ND_j$ and $PA_j$ the sets of parent nodes  and non-descendant nodes of $x_j$, respectively. 
Using decomposability  of $R$ with respect to the two ancestral sets
$ND_j\cup PA_j$ and $ND_j\cup PA_j \cup \{x_j\}$, we conclude
$$
I(nd_j:x_j|pa_j) = \sum_{x_a\in ND_j\cup PA_j} R(x_a|pa_a) + R(x_j|pa_j) - \sum_{x_a\in ND_j\cup PA_j\cup \{x_j\}} R(x_a|pa_a) = 0.
$$
$(3)\rightarrow (1)$ holds because the non-descendants of a node are $d$-separated from the node itself by the parents.\\
$(1)\rightarrow (3)$. Since the dependence measure $I$ satisfies the graphoid axioms (Lemma \ref{lemGraphoid}) we can apply Theorem 2 in Verma \& Pearl \cite{vermaPearl90} which asserts that the DAG is an $I$-map, or in other words that $d$-separation relations represent a subset of the (conditional) independences that hold for the given objects. $\Box$

\section{Proof of Theorem \ref{thmFaith}}\label{appFaith}

We only state the prove for the monotonely faithful case, since the proof in the faithful case is similar and has already been given in \cite{spirtes91}. The basic ingredient is Lemma A5 in \cite{spirtes91}, stating that 
\begin{enumerate}
\item[(*)] any two nodes $a$ and $b$ that are not adjacent in a DAG can be $d$-separated by a set of nodes $S_{ab}$ not containing $a$ and $b$.
\end{enumerate}
Let  the observations $X$ be faithfully represented by a graph $G_f$. We show first that $G_f$ fulfills $(1)$ and $(2)$: Since $d$-separation implies conditional independence in any causal model (Theorem \ref{thmMarkovEquiv}),  $(*)$ proves one direction of $(1)$. The converse direction follows because if $a$ and $b$ could be made  independent by conditioning on the join of nodes in a set $S_{ab}$, then there exists a minimal set $S'_{ab}$ with this property. By definition of a monotonely faithful representation,  $S'_{ab}$ $d$-separates $a$ and $b$, hence they can not be adjacent. To see $(2)$ let $a,b,c$ be nodes with the required adjacencies and assume $a\rightarrow b \leftarrow c$ in $G_f$. Since $a$ and $c$ are not adjacent, by $(*)$ there exists a minimal set $U$ that d-separates $a$ and $c$. Now $b$ is not a member of  $U$ and $d$-separation implies the desired independence. Conversely, assume there exists a set $U$ such that $a\independent c|u$, where $u$ is the join of the nodes in $U$. We can choose $U$ to be minimal with this property, thus monotone faithfulness of $G_f$ implies that $U$ $d$-separates $a$ and $c$. Because $b\notin U$ by assumption, the graphs $a \rightarrow b \rightarrow c$ and $a \leftarrow b\leftarrow x_c$ cannot be subgraphs of $G_f$.\\
Now let $G$ be a DAG that fulfills $(1)$ and $(2)$. We need to show that $G$ is a monotonely faithful representation of the observations in $X$. By assumption there exists a monotonely faithful DAG $G_f$ that, as we have seen, also fulfills $(1)$ and $(2)$. In particular, $(1)$ implies that $G$ and $G_f$ do have the same adjacencies and $(2)$ implies that they have the same subgraphs of the form $a\rightarrow b\leftarrow c$ $(2)$. Then a graph-theoretical argument (Lemma $A.8$ in \cite{spirtes91}) states that $G$ and $G_f$ imply the same $d$-separation relations, hence by definition $G$ is a monotonely faithfully representation, too.

\newpage
\bibliographystyle{unsrt}

\bibliography{literatur-causality}

\end{document}